\documentclass[runningheads]{llncs}

\usepackage[utf8]{inputenc}
\usepackage{centernot}
\usepackage{amssymb}
\usepackage[normalem]{ulem}
\usepackage{centernot}
\usepackage{xspace}
\usepackage[bookmarks,unicode,colorlinks=true]{hyperref}
\usepackage{stmaryrd}
\usepackage{mathrsfs}
\usepackage{proof}
\usepackage{bm}
\usepackage{tikz}
\usepackage{multirow}
\usepackage{array}
\usepackage{mathtools}
\usepackage{enumitem}
\usepackage{cleveref}
\usepackage{microtype}
\usepackage{cite} % sort references: [2,3,1] -> [1,2,3]

\usepackage[location=inline]{moveproofs}
\newcommand{\submission}[1]{}
\newcommand{\report}[1]{#1}

\submission{
  \usepackage{graphicx}
  \usepackage[firstpage]{draftwatermark}
  \SetWatermarkText{\hspace*{10cm}\raisebox{15.5cm}{\includegraphics{tacas-2020-ae-badge}}}
  \SetWatermarkAngle{0}
}

\makeatletter
\RequirePackage[bookmarks,unicode,colorlinks=true]{hyperref}%
   \def\@citecolor{blue}%
   \def\@urlcolor{blue}%
   \def\@linkcolor{blue}%

\def\orcidID#1{\smash{\href{http://orcid.org/#1}{\protect\raisebox{-1.25pt}{\protect\includegraphics{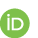}}}}}
\makeatother

\DeclareFontFamily{U}{mathx}{\hyphenchar\font45}
\DeclareFontShape{U}{mathx}{m}{n}{
      <5> <6> <7> <8> <9> <10>
      <10.95> <12> <14.4> <17.28> <20.74> <24.88>
      mathx10
      }{}
\DeclareSymbolFont{mathx}{U}{mathx}{m}{n}
\DeclareFontSubstitution{U}{mathx}{m}{n}
\DeclareMathAccent{\widecheck}{0}{mathx}{"71}

\hypersetup{
  pdftitle={A Calculus for Modular Loop Acceleration},
  pdfauthor={Florian Frohn}
}

\renewcommand{\epsilon}{\varepsilon}
\let\oldphi\phi
\let\oldvarphi\varphi
\renewcommand{\phi}{\oldvarphi}
\renewcommand{\varphi}{\oldphi}
\renewcommand{\hat}[1]{\widehat{#1}}
\renewcommand{\check}[1]{\widecheck{#1}}

\newcommand{\ZZ}{\mathbb{Z}}
\newcommand{\AAA}{\mathscr{C}}
\newcommand{\QQ}{\mathbb{Q}}
\newcommand{\TT}{\mathcal{T}}

\newcommand{\NN}{\mathbb{N}}
\newcommand{\charfun}[1]{I_{#1}}
\newcommand{\Def}{\mathrel{\mathop:}=}
\newcommand{\assign}{\leftarrow}

\newcommand{\mDo}{\mathbf{do}}
\newcommand{\mWhile}[2]{\mathbf{while}\ #1\ \mDo\ #2}
\newcommand{\relmiddle}[1]{\mathrel{}\middle#1\mathrel{}}
\newcommand{\Loop}{\mathit{Loop}}
\newcommand{\Prop}[1]{\mathit{Prop}(#1)}
\newcommand{\accelerate}{\mathit{accel}}
\newcommand{\mf}{\mathit{mf}}
\newcommand{\expr}{\mathit{expr}}
\newcommand{\mat}[1]{\left(\begin{smallmatrix} #1 \end{smallmatrix}\right)}
\newcommand{\tool}[1]{\textsf{#1}}
\newcommand{\loat}{\tool{LoAT}\xspace}
\newcommand{\prob}[4]{\left\llbracket #1 \relmiddle{|} #2 \relmiddle{|} #3 \relmiddle{|} #4 \right\rrbracket}

\DeclareMathOperator{\dom}{dom}
\DeclareMathOperator{\true}{\top}

\crefname{equation}{eq.}{equqations}%
\crefname{chapter}{chapter}{chapters}%
\crefname{section}{sec.}{sections}%
\crefname{appendix}{app.}{appendices}%
\crefname{enumi}{item}{items}%
\crefname{footnote}{footnote}{footnotes}%
\crefname{figure}{fig.}{figures}%
\crefname{table}{table}{tables}%
\crefname{theorem}{thm.}{theorems}%
\crefname{lemma}{lem.}{lemmas}%
\crefname{corollary}{cor.}{corollaries}%
\crefname{proposition}{proposition}{propositions}%
\crefname{definition}{def.}{definitions}%
\crefname{result}{result}{results}%
\crefname{example}{ex.}{examples}%
\crefname{remark}{remark}{remarks}%
\crefname{note}{note}{notes}%

\title{A Calculus for Modular Loop Acceleration\thanks{This work has been funded by DFG grant 389792660 as part of TRR~248 (see \url{https://perspicuous-computing.science}).}}
\author{Florian Frohn \orcidID{0000-0003-0902-1994}}
\institute{Max Planck Institute for Informatics\\Saarland Informatics Campus, Saarbr\"ucken, Germany}
\authorrunning{F.\ Frohn}

\begin{document}

\maketitle

\begin{abstract}
  Loop acceleration can be used to prove safety, reachability, runtime bounds, and (non-)termination of programs operating on integers.
  To this end, a variety of acceleration techniques has been proposed.
  However, all of them are monolithic:
  Either they accelerate a loop successfully or they fail completely.
  In contrast, we present a calculus that allows for combining acceleration techniques in a modular way and we show how to integrate many existing acceleration techniques into our calculus.
  Moreover, we propose two novel acceleration techniques that can be incorporated into our calculus seamlessly.
  An empirical evaluation demonstrates the applicability of our approach.
\end{abstract}
\section{Introduction}
\label{sec:intro}

In the last years, loop acceleration techniques have successfully been used to build static analyses for programs operating on integers \cite{underapprox15,ijcar16,fmcad19,journal,iosif17,Bozga14,fast}.
Essentially, such techniques extract a quantifier-free first-order formula $\psi$ from a single-path loop $\TT$, i.e., a loop without branching in its body, such that $\psi$ under-approximates (resp.\ is equivalent to) $\TT$.
More specifically, each model of the resulting formula $\psi$ corresponds to an execution of $\TT$ (and vice versa).
By integrating such techniques into a suitable program-analysis framework \cite{ijcar16,journal,fmcad19,iosif12,iosif17,FlatFramework}, whole programs can be transformed into first-order formulas which can then be analyzed by off-the-shelf solvers.
Applications include proving safety \cite{iosif12} or reachability \cite{iosif12,underapprox15}, deducing bounds on the runtime complexity \cite{ijcar16,journal}, and proving (non-) termination \cite{fmcad19,Bozga14}.

However, existing acceleration techniques only apply if certain prerequisites are in place.
So the power of static analyses built upon loop acceleration depends on the applicability of the underlying acceleration technique.

In this paper, we introduce a calculus which allows for combining several acceleration techniques modularly in order to accelerate a single loop.
Consequently, it can handle classes of loops where all standalone techniques fail.
Moreover, we present two novel acceleration techniques and integrate them into our calculus.

In the following, we introduce preliminaries in \Cref{sec:preliminaries}.
Then, we discuss existing acceleration techniques in \Cref{sec:monotonic}.
In \Cref{sec:integration}, we present our calculus to combine acceleration techniques.
\Cref{sec:conditional} shows how existing acceleration techniques can be integrated into our framework.
Next, we present two novel acceleration techniques and incorporate them into our calculus in \Cref{sec:accel}.
After discussing related work in \Cref{sec:related}, we demonstrate the applicability of our approach via an empirical evaluation in \Cref{sec:experiments} and conclude in \Cref{sec:conclusion}.
\submission{All proofs can be found in \cite{preprint}.}
\section{Preliminaries}
\label{sec:preliminaries}

We use bold letters $\vec{x}$, $\vec{y}$, $\vec{z}$, ... for vectors.
Let $\AAA(\vec{z})$ be the set of \emph{closed-form expressions} over the variables $\vec{z}$ containing, e.g., all arithmetic expressions built from $\vec{z}$, integer constants, addition, subtraction, multiplication, division, and exponentiation.\footnote{Note that there is no widely accepted definition of ``closed forms'' and the results of the current paper are independent of the precise definition of $\AAA(\vec{z})$.}
We consider loops of the form
\begin{equation}
  \label{loop}\tag{\ensuremath{\TT_{loop}}}
  \mWhile{\phi}{\vec{x} \assign \vec{a}}
\end{equation}
where $\vec{x}$ is a vector of $d$ pairwise different variables that range over the integers, the loop condition $\phi \in \Prop{\AAA(\vec{x})}$ is a finite propositional formula over the atoms $\{p>0 \mid p \in \AAA(\vec{x})\}$, and $\vec{a} \in \AAA(\vec{x})^d$ such that the function\footnote{i.e., the (anonymous) function that maps $\vec{x}$ to $\vec{a}$} $\vec{x} \mapsto \vec{a}$ maps integers to integers.
$\Loop$ denotes the set of all such loops.

We identify \ref{loop} and the pair $\langle \phi, \vec{a} \rangle$.
Moreover, we identify $\vec{a}$ and the function $\vec{x} \mapsto \vec{a}$ where we sometimes write $\vec{a}(\vec{x})$ to make the variables $\vec{x}$ explicit and we use the same convention for other (vectors of) expressions.
Similarly, we identify the formula $\phi$ resp.\ $\phi(\vec{x})$ and the predicate $\vec{x} \mapsto \phi$.

Throughout this paper, let $n$ be a designated variable and let:
\[
  \vec{a} \Def \mat{a_1\\\ldots\\a_d} \qquad \vec{x} \Def \mat{x_1\\\ldots\\x_d} \qquad \vec{x}' \Def \mat{x'_1\\\ldots\\x'_d} \qquad \vec{y} \Def \mat{\vec{x}\\n\\\vec{x}'}
\]
Intuitively, the variable $n$ represents the number of loop iterations and $\vec{x}'$ corresponds to the values of the program variables $\vec{x}$ after $n$ iterations.

$\ref{loop}$ induces a relation ${\longrightarrow_{\ref{loop}}}$ on $\ZZ^d$:
\[
  \phi(\vec{x}) \land \vec{x}' = \vec{a}(\vec{x}) \iff \vec{x} \longrightarrow_{\ref{loop}} \vec{x}'
\]
Our goal is to find a formula $\psi \in \Prop{\AAA(\vec{y})}$ such that
\begin{equation}
  \label{eq:equiv}\tag{equiv}
  \psi \iff \vec{x} \longrightarrow_{\ref{loop}}^n \vec{x}' \qquad \text{for all } n>0.
\end{equation}
To see why we use $\AAA(\vec{y})$ instead of, e.g., polynomials, consider the loop
\begin{equation}
  \label{loop:exp}
  \tag{\ensuremath{\TT_{exp}}}
  \mWhile{x_1>0}{\mat{x_1\\x_2} \assign \mat{x_1-1\\2 \cdot x_2}.}
\end{equation}
Here, an acceleration technique synthesizes, e.g., the formula
\begin{equation}
  \label{psi:exp}
  \tag{\ensuremath{\psi_{exp}}}
  \mat{x_1'\\x_2'} = \mat{x_1-n\\2^n \cdot x_2} \land x_1 - n + 1 > 0
\end{equation}
where $\mat{x_1-n\\2^n \cdot x_2}$ is equivalent to the value of $\mat{x_1\\x_2}$ after $n$ iterations and the inequation $x_1 - n + 1 > 0$ ensures that \ref{loop:exp} can be executed at least $n$ times.
Clearly, the growth of $x_2$ cannot be captured by a polynomial, i.e., even the behavior of quite simple loops is beyond the expressiveness of polynomial arithmetic.

In practice, one can restrict our approach to weaker classes of expressions to ease automation, but the presented results are independent of such considerations.

Some acceleration techniques cannot guarantee \eqref{eq:equiv}, but the resulting formula is an under-approximation of \ref{loop}, i.e., we have
\begin{equation}
  \label{eq:approx}\tag{approx}
  \psi \implies \vec{x} \longrightarrow_{\ref{loop}}^n \vec{x}' \qquad \text{for all } n>0.
\end{equation}
If \eqref{eq:equiv} resp.\ \eqref{eq:approx} holds, then $\psi$ is \emph{equivalent} to resp.\ \emph{approximates} \ref{loop}.

\begin{definition}[Acceleration Technique]
  An \emph{acceleration technique} is a partial function
  \[
    \accelerate: \Loop \rightharpoonup \Prop{\AAA(\vec{y})}.
  \]
  It is \emph{sound} if $\accelerate(\TT)$ approximates $\TT$ for all $\TT \in \dom(\accelerate)$.
  It is \emph{exact} if $\accelerate(\TT)$ is equivalent to $\TT$ for all $\TT \in \dom(\accelerate)$.
\end{definition}
\section{Existing Acceleration Techniques}
\label{sec:monotonic}

We now recall several existing acceleration techniques.
In \Cref{sec:integration} we will see how these techniques can be combined in a modular way.
All of them first compute a \emph{closed form} $\vec{c} \in \AAA(\vec{x},n)^d$ for the values of the program variables after $n$ iterations.
\begin{definition}[Closed Form]
  \label{def:closed}
  We call $\vec{c} \in \AAA(\vec{x},n)^d$ a \emph{closed form} of \ref{loop} if
  \[
    \forall \vec{x} \in \ZZ^d, n \in \NN.\ \vec{c} = \vec{a}^n(\vec{x}).
  \]
\end{definition}

Here, $\vec{a}^n$ is the $n$-fold application of $\vec{a}$, i.e., $\vec{a}^0(\vec{x}) = \vec{x}$ and $\vec{a}^{n+1}(\vec{x}) = \vec{a}(\vec{a}^n(\vec{x}))$.
To find closed forms, one tries to solve the system of recurrence equations $\vec{x}^{(n)} = \vec{a}(\vec{x}^{(n-1)})$ with the initial condition $\vec{x}^{(0)} = \vec{x}$.
In the sequel, we assume that we can represent $\vec{a}^n(\vec{x})$ in closed form.
Note that one can always do so if $\vec{a}(\vec{x}) = A\vec{x} + \vec{b}$ with $A \in \ZZ^{d \times d}$ and $\vec{b} \in \ZZ^d$, i.e., if $\vec{a}$ is affine.
To this end, one considers the matrix $B \Def \mat{A & \vec{b} \\ \vec{0}^T & 1}$ and computes its Jordan normal form $B = T^{-1}JT$ where $J$ is a block diagonal matrix (which has complex entries if $B$ has complex eigenvalues).
Then the closed form for $J^n$ can be given directly (see, e.g., \cite{Ouaknine15}) and $\vec{a}^n(\vec{x}) = T^{-1}J^nT\mat{\vec{x}\\1}$.
Moreover, one can compute a closed form if $\vec{a} = \mat{c_1 \cdot x_1 + p_1\\\ldots\\c_d \cdot x_d + p_d}$ where $c_i \in \NN$ and each $p_i$ is a polynomial over $x_1,\ldots,x_{i-1}$ \cite{polyloops}.

\subsection{Acceleration via Decrease \emph{or} Increase}
\label{subsec:kroening}

The first acceleration technique discussed in this section exploits the following observation:
If $\phi(\vec{a}(\vec{x}))$ implies $\phi(\vec{x})$ and $\phi(\vec{a}^{n-1}(\vec{x}))$ holds, then \ref{loop} is applicable at least $n$ times.
So in other words, it requires that the indicator function (or characteristic function) $\charfun{\phi}: \ZZ^d \to \{0,1\}$ of $\phi$ with $\charfun{\phi}(\vec{x}) = 1 \iff \phi(\vec{x})$ is monotonically decreasing w.r.t.\ $\vec{a}$, i.e., $\charfun{\phi}(\vec{x}) \geq \charfun{\phi}(\vec{a}(\vec{x}))$.
\begin{theorem}[Acceleration via Monotonic Decrease \cite{underapprox15}]
  \label{thm:one-way}
  If
  \[
    \phi(\vec{a}(\vec{x})) \implies \phi(\vec{x}),
  \]
  then the following acceleration technique is exact:
  \[
    \ref{loop} \mapsto \vec{x}' = \vec{a}^n(\vec{x}) \land \phi(\vec{a}^{n-1}(\vec{x}))
  \]
\end{theorem}
\makeproof{thm:one-way}{We will prove the more general \Cref{thm:conditional-one-way} in \Cref{sec:conditional}. \qed}

So for example, \Cref{thm:one-way} accelerates \ref{loop:exp} to \ref{psi:exp}.
However, the requirement $\phi(\vec{a}(\vec{x})) \implies \phi(\vec{x})$ is often violated in practice.
To see this, consider the loop
\begin{equation}
  \label{eq:conditional-ex}\tag{\ensuremath{\TT_{non\text{-}dec}}}
  \mWhile{x_1 > 0 \land x_2 > 0}{\mat{x_1\\x_2} \assign \mat{x_1 - 1\\x_2 + 1}}.
\end{equation}
It cannot be accelerated with \Cref{thm:one-way} as
\[
  x_1-1 > 0 \land x_2+1 > 0 \centernot\implies x_1 > 0 \land x_2 > 0.
\]

A dual acceleration technique is obtained by ``reversing'' the implication in the prerequisites of \Cref{thm:one-way}.
Then $\charfun{\phi}$ is monotonically increasing w.r.t.\ $\vec{a}$.
So $\phi$ is an invariant and thus $\{\vec{x} \in \ZZ^d \mid \phi(\vec{x})\}$ is a \emph{recurrent set} \cite{rupak08} of \ref{loop}.

\begin{theorem}[Acceleration via Monotonic Increase]
  \label{thm:recurrent}
  If
  \[
    \phi(\vec{x}) \implies \phi(\vec{a}(\vec{x})),
  \]
  then the following acceleration technique is exact:
  \[
    \ref{loop} \mapsto \vec{x}' = \vec{a}^n(\vec{x}) \land \phi(\vec{x})
  \]
\end{theorem}
\makeproof{thm:recurrent}{We will prove the more general \Cref{thm:conditional-recurrent} in \Cref{sec:conditional}. \qed}

As a minimal example, \Cref{thm:recurrent} accelerates
\[
  \mWhile{x > 0}{x \assign x+1}
\]
to $x' = x + n \land x > 0$.

\subsection{Acceleration via Decrease \emph{and} Increase}
\label{subsec:three-way}

Both acceleration techniques presented so far have been generalized in \cite{fmcad19}.
\begin{theorem}[Acceleration via Monotonicity \cite{fmcad19}]
  \label{thm:three-way}
  If
  \begin{align*}
    \phi(\vec{x}) \iff& \phi_1(\vec{x}) \land \phi_2(\vec{x}) \land \phi_3(\vec{x}),\\
    \phi_1(\vec{x}) \implies& \phi_1(\vec{a}(\vec{x})),\\
    \phi_1(\vec{x}) \land \phi_2(\vec{a}(\vec{x})) \implies& \phi_2(\vec{x}), & \text{and} \\
    \phi_1(\vec{x}) \land \phi_2(\vec{x}) \land \phi_3(\vec{x}) \implies& \phi_3(\vec{a}(\vec{x})),
  \end{align*}
  then the following acceleration technique is exact:
  \[
    \ref{loop} \mapsto \vec{x}' = \vec{a}^n(\vec{x}) \land \phi_1(\vec{x}) \land \phi_2(\vec{a}^{n-1}(\vec{x})) \land \phi_3(\vec{x})
  \]
\end{theorem}
\makeproof{thm:three-way}{Immediate consequence of \Cref{cor:calculus-sound} and \Cref{thm:simulate-three-way}, which will be proven in \Cref{sec:integration,sec:conditional}. \qed}

Here, $\phi_1$ and $\phi_3$ are again invariants of the loop.
Thus, as in \Cref{thm:recurrent} it suffices to require that they hold before entering the loop.
On the other hand, $\phi_2$ needs to satisfy a similar condition as in \Cref{thm:one-way} and thus it suffices to require that $\phi_2$ holds before the last iteration.
We also say that $\phi_2$ is a \emph{converse invariant} (w.r.t.\ $\phi_1$).
It is easy to see that \Cref{thm:three-way} is equivalent to \Cref{thm:one-way} if $\phi_1 \equiv \phi_3 \equiv \true$ (where $\true$ denotes logical truth) and it is equivalent to \Cref{thm:recurrent} if $\phi_2 \equiv \phi_3 \equiv \true$.

With this approach, \ref{eq:conditional-ex} can be accelerated to
\begin{equation}
  \label{psi:conditional-ex}
  \tag{\ensuremath{\psi_{non\text{-}dec}}}
  \mat{x_1'\\x_2'} = \mat{x_1-n\\x_2+n} \land x_2 > 0 \land x_1-n+1 > 0
\end{equation}
by choosing $\phi_1 \Def x_2 > 0$, $\phi_2 \Def x_1 > 0$, and $\phi_3 \Def \true$.

\Cref{thm:three-way} naturally raises the question: Why do we need \emph{two} invariants?
To see this, consider a restriction of \Cref{thm:three-way} where $\phi_3 \Def \true$.
It would fail for a loop like
\begin{equation}
  \label{eq:three-way-ex}
  \tag{\ensuremath{\TT_{2\text{-}invs}}}
  \mWhile{x_1 > 0 \land x_2 > 0}{\mat{x_1\\x_2} \assign \mat{x_1+x_2\\x_2-1}}
\end{equation}
which can easily be handled by \Cref{thm:three-way} (by choosing $\phi_1 \Def \true$, $\phi_2 \Def x_2 > 0$, and $\phi_3 \Def x_1 > 0$).
The problem is that the converse invariant $x_2 > 0$ is needed to prove invariance of $x_1 > 0$.
Similarly, a restriction of \Cref{thm:three-way} where $\phi_1 \Def \true$ would fail for the following variant of \ref{eq:three-way-ex}:
\[
  \mWhile{x_1 > 0 \land x_2 > 0}{\mat{x_1\\x_2} \assign \mat{x_1-x_2\\x_2+1}}
\]
Here, the problem is that the invariant $x_2 > 0$ is needed to prove converse invariance of $x_1 > 0$.

\subsection{Acceleration via Metering Functions}
\label{sec:metering}

Another approach for loop acceleration uses \emph{metering functions}, a variation of classical \emph{ranking functions} from termination and complexity analysis \cite{ijcar16}.
While ranking functions give rise to \emph{upper} bounds on the runtime of loops, metering functions provide \emph{lower} runtime bounds, i.e., the definition of a metering function $\mf: \ZZ^d \to \QQ$ ensures that for each $\vec{x} \in \ZZ^d$, the loop under consideration can be applied at least $\lceil \mf(\vec{x}) \rceil$ times.

\begin{theorem}[Acceleration via Metering Functions \cite{ijcar16}]
  \label{thm:meter}
  Let $\mf$ be a metering function for \ref{loop}.
  Then the following acceleration technique is sound:
  \[
    \ref{loop} \mapsto \vec{x}' = \vec{a}^n(\vec{x}) \land \phi(\vec{x}) \land n < \mf(\vec{x}) + 1
  \]
\end{theorem}
\makeproof{thm:meter}{We will prove the more general \Cref{thm:conditional-metering} in \Cref{sec:conditional}. \qed}

So using the metering function $x$, \Cref{thm:meter} accelerates \ref{loop:exp} to
\[
  \mat{x_1'\\x_2'} = \mat{x_1-n\\2^n \cdot x_2} \land x_1 > 0 \land n < x_1 + 1 \quad \equiv \quad \ref{psi:exp}.
\]

However, synthesizing non-trivial (i.e., non-constant) metering functions is challenging.
Moreover, unless the number of iterations of \ref{loop} equals $\lceil \mf(\vec{x}) \rceil$ for all $\vec{x} \in \ZZ^d$, \emph{acceleration via metering functions} is not exact.

\emph{Linear} metering functions can be synthesized via Farkas' Lemma and SMT solving \cite{ijcar16}.
However, many loops do not have non-trivial linear metering functions.
To see this, reconsider \ref{eq:conditional-ex}.
Here, $(x_1,x_2) \mapsto x_1$ is not a metering function as \ref{eq:conditional-ex} cannot be iterated at least $x_1$ times if $x_2 \leq 0$.
Thus, \cite{journal} proposes a refinement of \cite{ijcar16} based on metering functions of the form $\vec{x} \mapsto \charfun{\xi}(\vec{x}) \cdot f(\vec{x})$ where $\xi \in \Prop{\AAA(\vec{x})}$ and $f$ is linear.
With this improvement, the metering function $(x_1,x_2) \mapsto \charfun{x_2 > 0}(x_2) \cdot x_1$ can be used to accelerate \ref{eq:conditional-ex} to
\[
  \mat{x_1'\\x_2'} = \mat{x_1-n\\x_2+n} \land x_1 > 0 \land x_2 > 0 \land n < x_1 + 1.
\]

\section{A Calculus for Modular Loop Acceleration}
\label{sec:integration}

All acceleration techniques presented so far are monolithic:
Either they accelerate a loop successfully or they fail completely.
In other words, we cannot \emph{combine} several techniques to accelerate a single loop.
To this end, we now present a calculus that repeatedly applies acceleration techniques to simplify an \emph{acceleration problem} resulting from a loop \ref{loop} until it is \emph{solved} and hence gives rise to a suitable $\psi \in \Prop{\AAA(\vec{y})}$ which approximates resp.\ is equivalent to \ref{loop}.
\begin{definition}[Acceleration Problem]
  A tuple
  \[
    \prob{\psi}{\check{\phi}}{\hat{\phi}}{\vec{a}}
  \]
  where $\psi \in \Prop{\AAA(\vec{y})}$, $\check{\phi}, \hat{\phi} \in \Prop{\AAA(\vec{x})}$, and $\vec{a}: \ZZ^d \to \ZZ^d$ is an \emph{acceleration problem}.
  It is \emph{consistent} if $\psi$ approximates $\langle \check{\phi}, \vec{a} \rangle$, \emph{exact} if $\psi$ is equivalent to $\langle \check{\phi}, \vec{a} \rangle$, and \emph{solved} if it is consistent and $\hat{\phi} \equiv \true$.
  The \emph{canonical acceleration problem} of a loop \ref{loop} is
  \[
    \prob{\vec{x}' = \vec{a}^n(\vec{x})}{\true}{\phi(\vec{x})}{\vec{a}(\vec{x})}.
  \]
\end{definition}
\begin{example}
  \label{ex:canonical}
  The canonical acceleration problem of \ref{eq:conditional-ex} is
  \[
    \prob{\mat{x_1'\\x_2'} = \mat{x_1-n\\x_2+n}}{\true}{x_1>0 \land x_2>0}{\mat{x_1-1\\x_2+1}}.
  \]
\end{example}

The first component $\psi$ of an acceleration problem $\prob{\psi}{\check{\phi}}{\hat{\phi}}{\vec{a}}$ is the partial result that has been computed so far.
The second component $\check{\phi}$ corresponds to the part of the loop condition that has already been processed successfully.
As our calculus preserves consistency, $\psi$ always approximates $\langle \check{\phi}, \vec{a} \rangle$.
The third component is the part of the loop condition that remains to be processed, i.e., the loop $\langle \hat{\phi},\vec{a} \rangle$ still needs to be accelerated.
The goal of our calculus is to transform a canonical into a solved acceleration problem.

More specifically, when we have simplified a canonical acceleration problem $\prob{\vec{x}' = \vec{a}^n(\vec{x})}{\true}{\phi(\vec{x})}{\vec{a}(\vec{x})}$ to $\prob{\psi_1(\vec{y})}{\check{\phi}(\vec{x})}{\hat{\phi}(\vec{x})}{\vec{a}(\vec{x})}$, then $\phi \equiv \check{\phi} \land \hat{\phi}$ and
\[
  \psi_1 \implies \vec{x} \longrightarrow^n_{\langle \check{\phi}, \vec{a} \rangle} \vec{x}'.
\]
Thus, it then suffices to find some $\psi_2 \in \Prop{\AAA(\vec{y})}$ such that
\begin{equation}
  \label{eq:post}
  \vec{x} \longrightarrow^n_{\langle \check{\phi}, \vec{a} \rangle} \vec{x}' \land \psi_2 \implies \vec{x} \longrightarrow^n_{\langle \hat{\phi}, \vec{a} \rangle} \vec{x}'.
\end{equation}
The reason is that we have ${\longrightarrow_{\langle \check{\phi}, \vec{a} \rangle}} \cap {\longrightarrow_{\langle \hat{\phi}, \vec{a} \rangle}} = {\longrightarrow_{\langle \check \phi \land \hat \phi, \vec{a} \rangle}} = {\longrightarrow_{\langle \phi, \vec{a} \rangle}}$ and thus
\[
  \psi_1 \land \psi_2 \implies \vec{x} \longrightarrow^n_{\langle \phi, \vec{a} \rangle} \vec{x}',
\]
i.e., $\psi_1 \land \psi_2$ approximates \ref{loop}.

Note that the acceleration techniques presented so far would map $\langle \hat{\phi}, \vec{a} \rangle$ to some $\psi_2 \in \Prop{\AAA(\vec{y})}$ such that
\begin{equation}
  \label{eq:current}
  \psi_2 \implies \vec{x} \longrightarrow^n_{\langle \hat{\phi}, \vec{a} \rangle} \vec{x}',
\end{equation}
which is more restrictive than \eqref{eq:post}.
In \Cref{sec:conditional}, we will adapt all acceleration techniques from \Cref{sec:monotonic} to search for some $\psi_2 \in \Prop{\AAA(\vec{y})}$ that satisfies \eqref{eq:post} instead of \eqref{eq:current}, i.e., we will turn them into \emph{conditional acceleration techniques}.
\begin{definition}[Conditional Acceleration]
  \label{def:cond-accel}
  We call a partial function
  \[
    \accelerate: \Loop \times \Prop{\AAA(\vec{x})} \rightharpoonup \Prop{\AAA(\vec{y})}.
  \]
  a \emph{conditional acceleration technique}.
  It is \emph{sound} if
  \[
    \vec{x} \longrightarrow^n_{\langle \check{\phi}, \vec{a} \rangle} \vec{x}' \land \accelerate(\langle \chi, \vec{a} \rangle,\check{\phi}) \quad \text{implies} \quad \vec{x} \longrightarrow^n_{\langle \chi, \vec{a} \rangle} \vec{x}'
  \]
  for all $(\langle \chi, \vec{a} \rangle,\check{\phi}) \in \dom(\accelerate)$, $\vec{x},\vec{x}' \in \ZZ^d$, and $n > 0$.
  It is \emph{exact} if additionally
  \[
    \vec{x} \longrightarrow^n_{\langle \chi \land \check{\phi}, \vec{a} \rangle} \vec{x}' \quad \text{implies} \quad \accelerate(\langle \chi, \vec{a} \rangle,\check{\phi})
  \]
  for all $(\langle \chi, \vec{a} \rangle,\check{\phi}) \in \dom(\accelerate)$, $\vec{x},\vec{x}' \in \ZZ^d$, and $n > 0$.
\end{definition}

We are now ready to present our \emph{acceleration calculus}, which combines loop acceleration techniques in a modular way.
In the following, w.l.o.g.\ we assume that propositional formulas are in CNF and we identify the formula $\bigwedge_{i=1}^k C_i$ with the set of clauses $\{C_i \mid 1 \leq i \leq k\}$.
\begin{definition}[Acceleration Calculus]
  \label{def:calculus}
  The relation ${\leadsto}$ on acceleration problems is defined by the following rule:
  \[
    \infer[
    \begin{array}{r}
      \accelerate \text { is a sound condition-}\\
      \text{al acceleration technique}
    \end{array}
    ]{
      \prob{\psi_1}{\check{\phi}}{\hat{\phi}}{\vec{a}} \leadsto_{(e)} \prob{\psi_1 \cup \psi_2}{\check{\phi} \cup \chi}{\hat{\phi} \setminus \chi}{\vec{a}}
    }{
      \emptyset \neq \chi \subseteq \hat{\phi} & \accelerate(\langle \chi, \vec{a} \rangle, \check{\phi}) = \psi_2
    }
  \]
  A ${\leadsto}$-step is \emph{exact} (written ${\leadsto_e}$) if $\accelerate$ is exact.
\end{definition}

So our calculus allows us to pick a subset $\chi$ (of clauses) from the yet unprocessed condition $\hat{\phi}$ and ``move'' it to $\check\phi$, which has already been processed successfully.
To this end, $\langle \chi, \vec{a} \rangle$ needs to be accelerated by a conditional acceleration technique, i.e., when accelerating $\langle \chi, \vec{a} \rangle$ we may assume $\vec{x} \longrightarrow_{\langle \check\phi, \vec{a} \rangle}^n \vec{x}'$.

Note that every acceleration technique trivially gives rise to a conditional acceleration technique (by disregarding the second argument $\check{\phi}$ of $\accelerate$ in \Cref{def:cond-accel}).
Thus, our calculus allows for combining arbitrary existing acceleration techniques without adapting them.
However, many acceleration techniques can easily be turned into more sophisticated conditional acceleration techniques (cf.\ \Cref{sec:conditional}), which increases the power of our approach.

\begin{example}
  \label{ex:calculus}
  We continue \Cref{ex:canonical} and fix $\chi \Def x_1>0$.
  Thus, we need to accelerate the loop $\left\langle x_1>0, \mat{x_1-1\\x_2+1} \right\rangle$ to enable a $\leadsto$-step.
  We obtain
  \[
    \begin{array}{cl}
      & \prob{\psi^{init}_{non\text{-}dec} \Def \mat{x_1'\\x_2'} = \mat{x_1-n\\x_2+n}}{\true}{x_1>0 \land x_2>0}{\mat{x_1-1\\x_2+1}} \\[.5em]
      \overset{\Cref{thm:one-way}}{\leadsto_e} & \prob{\psi^{init}_{non\text{-}dec} \land x_1-n+1 > 0}{x_1>0}{x_2>0}{\mat{x_1-1\\x_2+1}} \\[.5em]
      \overset{\Cref{thm:recurrent}}{\leadsto_e} & \prob{\psi^{init}_{non\text{-}dec} \land x_1-n+1 > 0 \land x_2 > 0}{x_1>0 \land x_2>0}{\true}{\mat{x_1-1\\x_2+1}} \\[.5em]
      = & \prob{\ref{psi:conditional-ex}}{x_1 > 0 \land x_2 > 0}{\true}{\mat{x_1-1\\x_2+1}}
    \end{array}
  \]
  where \Cref{thm:recurrent} was applied to the loop $\left\langle x_2>0, \mat{x_1-1\\x_2+1} \right\rangle$ in the second step.
  Thus, we successfully constructed the formula \ref{psi:conditional-ex}, which is equivalent to \ref{eq:conditional-ex}.
\end{example}

The crucial property of our calculus is the following.
\begin{lemma}
  \label{thm:calculus-sound}
  $\leadsto$ preserves consistency and $\leadsto_e$ preserves exactness.
\end{lemma}
\makeproof{thm:calculus-sound}{
  Assume
  \[
    \prob{\psi_1}{\check{\phi}}{\hat{\phi}}{\vec{a}} \leadsto \prob{\psi_1 \cup \psi_2}{\check{\phi} \cup \chi}{\hat{\phi} \setminus \chi}{\vec{a}}
  \]
  where $\prob{\psi_1}{\check{\phi}}{\hat{\phi}}{\vec{a}}$ is consistent and $\accelerate(\langle \chi, \vec{a} \rangle, \check{\phi}) = \psi_2$.
  We get
  \begin{align*}
    &\psi_1 \land \psi_2\\
    \implies& \vec{x} \longrightarrow^n_{\langle \check{\phi}, \vec{a} \rangle} \vec{x}' \land \psi_2 \tag{by consistency of $\prob{\psi_1}{\check{\phi}}{\hat{\phi}}{\vec{a}}$}\\
    \implies& \vec{x} \longrightarrow^n_{\langle \check{\phi}, \vec{a} \rangle} \vec{x}' \land \vec{x} \longrightarrow^n_{\langle \chi, \vec{a} \rangle} \vec{x} \tag{by soundness of $\accelerate$}\\
    \iff& \vec{x} \longrightarrow^n_{\langle \check{\phi} \land \chi, \vec{a} \rangle} \vec{x}'
  \end{align*}
  This proves consistency of
  \[
    \prob{\psi_1 \land \psi_2}{\check{\phi} \land \chi}{\hat{\phi} \setminus \chi}{\vec{a}} =
    \prob{\psi_1 \cup \psi_2}{\check{\phi} \cup \chi}{\hat{\phi} \setminus \chi}{\vec{a}}.
  \]

  Now assume
  \[
    \prob{\psi_1}{\check{\phi}}{\hat{\phi}}{\vec{a}} \leadsto_e \prob{\psi_1 \cup \psi_2}{\check{\phi} \cup \chi}{\hat{\phi} \setminus \chi}{\vec{a}}
  \]
  where $\prob{\psi_1}{\check{\phi}}{\hat{\phi}}{\vec{a}}$ is exact and $\accelerate(\langle \chi, \vec{a} \rangle, \check{\phi}) = \psi_2$.
  We get
  \begin{align*}
    &\vec{x} \longrightarrow^n_{\langle \check{\phi} \land \chi, \vec{a} \rangle} \vec{x}'\\
    \implies & \vec{x} \longrightarrow^n_{\langle \check{\phi} \land \chi, \vec{a} \rangle} \vec{x}' \land \psi_2 \tag{by exactness of $\accelerate$}\\
    \implies & \vec{x} \longrightarrow^n_{\langle \check{\phi}, \vec{a} \rangle} \vec{x}' \land \psi_2\\
    \iff & \psi_1 \land \psi_2 \tag{by exactness of $\prob{\psi_1}{\check{\phi}}{\hat{\phi}}{\vec{a}}$}\\
  \end{align*}
  which, together with consistency, proves exactness of
  \[
     \prob{\psi_1 \land \psi_2}{\check{\phi} \land \chi}{\hat{\phi} \setminus \chi}{\vec{a}} = \prob{\psi_1 \cup \psi_2}{\check{\phi} \cup \chi}{\hat{\phi} \setminus \chi}{\vec{a}}.
  \]\qed
}

Then the correctness of our calculus follows immediately.
The reason is that $\prob{\vec{x}' = \vec{a}^n(\vec{x})}{\true}{\phi(\vec{x})}{\vec{a}(\vec{x})} \leadsto_{(e)}^* \prob{\psi(\vec{y})}{\check{\phi}(\vec{x})}{\true}{\vec{a}(\vec{x})}$ implies $\phi \equiv \check{\phi}$.

\begin{theorem}[Correctness of ${\leadsto}$]
  \label{cor:calculus-sound}
  If
  \[
    \prob{\vec{x}' = \vec{a}^n(\vec{x})}{\true}{\phi(\vec{x})}{\vec{a}(\vec{x})} \leadsto^* \prob{\psi(\vec{y})}{\check{\phi}(\vec{x})}{\true}{\vec{a}(\vec{x})},
  \]
  then $\psi$ approximates \ref{loop}.
  If
  \[
    \prob{\vec{x}' = \vec{a}^n(\vec{x})}{\true}{\phi(\vec{x})}{\vec{a}(\vec{x})} \leadsto_e^* \prob{\psi(\vec{y})}{\check{\phi}(\vec{x})}{\true}{\vec{a}(\vec{x})},
  \]
  then $\psi$ is equivalent to \ref{loop}.
\end{theorem}

Termination of our calculus is trivial, as the size of the third component $\hat{\phi}$ of the acceleration problem is decreasing.

\begin{theorem}[Termination of ${\leadsto}$]
  \label{thm:term}
  $\leadsto$ terminates.
\end{theorem}

\section{Conditional Acceleration Techniques}
\label{sec:conditional}

We now show how to turn the acceleration techniques from \Cref{sec:monotonic} into conditional acceleration techniques, starting with \emph{acceleration via monotonic decrease}.
\begin{theorem}[Conditional Acceleration via Monotonic Decrease]
  \label{thm:conditional-one-way}
  If
  \begin{equation}
    \label{eq:conditional-one-way}
    \submission{\notag}
    \check{\phi}(\vec{x}) \land \chi(\vec{a}(\vec{x})) \implies \chi(\vec{x}),
  \end{equation}
  then the following conditional acceleration technique is exact:
  \[
    (\langle \chi, \vec{a} \rangle, \check{\phi}) \mapsto \vec{x}' = \vec{a}^n(\vec{x}) \land \chi(\vec{a}^{n-1}(\vec{x}))
  \]
\end{theorem}
\makeproof{thm:conditional-one-way}{
  For soundness, we need to prove
  \begin{equation}
    \label{thm:conditional-one-way-ih}
    \vec{x} \longrightarrow^m_{\langle \check{\phi}, \vec{a} \rangle} \vec{a}^{m}(\vec{x}) \land \chi(\vec{a}^{m-1}(\vec{x})) \implies \vec{x} \longrightarrow^m_{\langle \chi, \vec{a} \rangle} \vec{a}^{m}(\vec{x})
  \end{equation}
  for all $m > 0$.
  We use induction on $m$.
  If $m=1$, then
  \begin{align*}
    &\vec{x} \longrightarrow^m_{\langle \check{\phi}, \vec{a} \rangle} \vec{a}^{m}(\vec{x}) \land \chi(\vec{a}^{m-1}(\vec{x}))\\
    \implies & \chi(\vec{x}) \tag{as $m=1$}\\
    \implies & \vec{x} \longrightarrow_{\langle \chi, \vec{a} \rangle} \vec{a}(\vec{x})\\
    \iff & \vec{x} \longrightarrow^m_{\langle \chi, \vec{a} \rangle} \vec{a}^m(\vec{x}). \tag{as $m=1$}
  \end{align*}
  In the induction step, we have
  \begin{align*}
    &\vec{x} \longrightarrow^{m+1}_{\langle \check{\phi}, \vec{a} \rangle} \vec{a}^{m+1}(\vec{x}) \land \chi(\vec{a}^m(\vec{x})) \\
    \implies & \vec{x} \longrightarrow^{m}_{\langle \check{\phi}, \vec{a} \rangle} \vec{a}^{m}(\vec{x}) \land \chi(\vec{a}^m(\vec{x}))\\
    \implies & \vec{x} \longrightarrow^{m}_{\langle \check{\phi}, \vec{a} \rangle} \vec{a}^{m}(\vec{x}) \land \check{\phi}(\vec{a}^{m-1}(\vec{x})) \land \chi(\vec{a}^m(\vec{x})) \tag{as $m > 0$}\\
    \implies & \vec{x} \longrightarrow^{m}_{\langle \check{\phi}, \vec{a} \rangle} \vec{a}^{m}(\vec{x}) \land \chi(\vec{a}^{m-1}(\vec{x})) \land \chi(\vec{a}^m(\vec{x})) \tag{due to \eqref{eq:conditional-one-way}} \\
    \implies & \vec{x} \longrightarrow^m_{\langle \chi, \vec{a} \rangle} \vec{a}^m(\vec{x}) \land \chi(\vec{a}^m(\vec{x})) \tag{by the induction hypothesis \eqref{thm:conditional-one-way-ih}}\\
    \implies & \vec{x} \longrightarrow^{m+1}_{\langle \chi, \vec{a} \rangle} \vec{a}^{m+1}(\vec{x}).
  \end{align*}

  For exactness, we need to prove
  \[
    \vec{x} \longrightarrow^m_{\langle \chi \land \check{\phi}, \vec{a} \rangle} \vec{a}^m(\vec{x}) \implies \chi(\vec{a}^{m-1}(\vec{x}))
  \]
  for all $m>0$, which is trivial. \qed
}

So we just add $\check{\phi}$ to the premise of the implication that needs to be checked to apply \emph{acceleration via monotonic decrease}.
\Cref{thm:recurrent} can be adapted analogously.
\begin{theorem}[Conditional Acceleration via Monotonic Increase]
  \label{thm:conditional-recurrent}
  If
  \begin{equation}
    \label{eq:conditional-recurrent}
    \submission{\notag}
    \check{\phi}(\vec{x}) \land \chi(\vec{x}) \implies \chi(\vec{a}(\vec{x})),
  \end{equation}
  then the following conditional acceleration technique is exact:
  \[
    (\langle \chi, \vec{a} \rangle, \check{\phi}) \mapsto \vec{x}' = \vec{a}^n(\vec{x}) \land \chi(\vec{x})
  \]
\end{theorem}
\makeproof{thm:conditional-recurrent}{
  For soundness, we need to prove
  \begin{equation}
    \label{eq:conditional-recurrent-ih}
    \vec{x} \longrightarrow^m_{\langle \check\phi, \vec{a} \rangle} \vec{a}^m(\vec{x}) \land \chi(\vec{x}) \implies \vec{x} \longrightarrow^m_{\langle \chi, \vec{a} \rangle} \vec{a}^m(\vec{x})
  \end{equation}
  for all $m>0$.
  We use induction on $m$. If $m=1$, then
  \begin{align*}
    & \vec{x} \longrightarrow^m_{\langle \check\phi, \vec{a} \rangle} \vec{a}^m(\vec{x}) \land \chi(\vec{x})\\
    \implies & \vec{x} \longrightarrow_{\langle \chi, \vec{a} \rangle} \vec{a}(\vec{x}) \\
    \iff & \vec{x} \longrightarrow^m_{\langle \chi, \vec{a} \rangle} \vec{a}^m(\vec{x}).\tag{as $m=1$}
  \end{align*}
  In the induction step, we have
  \begin{align*}
    & \vec{x} \longrightarrow^{m+1}_{\langle \check\phi, \vec{a} \rangle} \vec{a}^{m+1}(\vec{x}) \land \chi(\vec{x})\\
    \implies & \vec{x} \longrightarrow^{m}_{\langle \check\phi, \vec{a} \rangle} \vec{a}^{m}(\vec{x}) \land \chi(\vec{x})\\
    \implies & \vec{x} \longrightarrow^{m}_{\langle \check\phi, \vec{a} \rangle} \vec{a}^{m}(\vec{x}) \land \vec{x} \longrightarrow^m_{\langle \chi, \vec{a} \rangle} \vec{a}^m(\vec{x}) \tag{by the induction hypothesis \eqref{eq:conditional-recurrent-ih}} \\
    \implies & \vec{x} \longrightarrow^m_{\langle \chi, \vec{a} \rangle} \vec{a}^m(\vec{x}) \land \check\phi(\vec{a}^{m-1}(\vec{x})) \land \chi(\vec{a}^{m-1}(\vec{x})) \tag{as $m > 0$}\\
    \implies & \vec{x} \longrightarrow^m_{\langle \chi, \vec{a} \rangle} \vec{a}^m(\vec{x}) \land \chi(\vec{a}^{m}(\vec{x})) \tag{due to \eqref{eq:conditional-recurrent}} \\
    \implies & \vec{x} \longrightarrow^{m+1}_{\langle \chi, \vec{a} \rangle} \vec{a}^{m+1}(\vec{x}).
  \end{align*}

  For exactness, we need to prove
  \[
    \vec{x} \longrightarrow^m_{\langle \chi \land \check\phi, \vec{a} \rangle} \vec{a}^m(\vec{x}) \implies \chi(\vec{x}),
  \]
  for all $m>0$, which is trivial. \qed
}

\begin{example}
  For the canonical acceleration problem of \ref{eq:three-way-ex}, we obtain:
  \begin{align*}
    & \prob{\vec{x}' = \vec{a}_{2\text{-}invs}^n(\vec{x})}{\true}{x_1>0 \land x_2>0}{\vec{a}_{2\text{-}invs} \Def \mat{x_1+x_2\\x_2-1}} \\
    \overset{\Cref{thm:conditional-one-way}}{\leadsto_e} & \prob{\vec{x}' = \vec{a}_{2\text{-}invs}^n(\vec{x}) \land x_2 - n + 1 > 0}{x_2>0}{x_1>0}{\vec{a}_{2\text{-}invs}} \\
    \overset{\Cref{thm:conditional-recurrent}}{\leadsto_e} & \prob{\vec{x}' = \vec{a}_{2\text{-}invs}^n(\vec{x}) \land x_2 - n + 1 > 0 \land x_1 > 0}{x_2>0 \land x_1>0}{\true}{\vec{a}_{2\text{-}invs}}
  \end{align*}
  While we could also use \Cref{thm:one-way} for the first step, \Cref{thm:recurrent} is inapplicable in the second step.
  The reason is that we need the converse invariant $x_2>0$ to prove invariance of $x_1 > 0$.
\end{example}

It is not a coincidence that \ref{eq:three-way-ex}, which could also be accelerated with \emph{acceleration via monotonicity} (cf.\ \Cref{thm:three-way}) directly, can be handled by applying our novel calculus with \Cref{thm:conditional-one-way,thm:conditional-recurrent}.
\begin{remark}
  \label{thm:simulate-three-way}
  If applying \emph{acceleration via monotonicity} to \ref{loop} yields $\psi$, then
  \[
    \prob{\vec{x}' = \vec{a}^n(\vec{x})}{\true}{\phi(\vec{x})}{\vec{a}(\vec{x})} \leadsto_e^{\leq 3} \prob{\psi(\vec{y})}{\phi(\vec{x})}{\true}{\vec{a}(\vec{x})}
  \]
  where either \Cref{thm:conditional-one-way} or \Cref{thm:conditional-recurrent} is applied in each ${\leadsto_e}$-step.
\end{remark}
\makeproof{thm:simulate-three-way}{
  As \Cref{thm:three-way} applies, we have $\phi(\vec{x}) \equiv \phi_1(\vec{x}) \land \phi_2(\vec{x}) \land \phi_3(\vec{x})$ where
  \begin{align}
    \phi_1(\vec{x}) &\implies \phi_1(\vec{a}(\vec{x})) & \land \label{eq:simulate-three-way1} \\
    \phi_1(\vec{x}) \land \phi_2(\vec{a}(\vec{x})) &\implies \phi_2(\vec{x}) & \land \label{eq:simulate-three-way2}\\
    \phi_1(\vec{x}) \land \phi_2(\vec{x}) \land \phi_3(\vec{x}) &\implies \phi_3(\vec{a}(\vec{x})). \label{eq:simulate-three-way3}
  \end{align}
  If $\phi_1 \neq \true$, then \Cref{thm:conditional-recurrent} applies to $\langle \phi_1, \vec{a} \rangle$ with $\check{\phi} \Def \true$ due to \eqref{eq:simulate-three-way1} and we obtain
  \begin{align*}
    & \prob{\vec{x}' = \vec{a}^n(\vec{x})}{\true}{\phi(\vec{x})}{\vec{a}(\vec{x})}\\
    = & \prob{\vec{x}' = \vec{a}^n(\vec{x})}{\true}{\phi_1(\vec{x}) \land \phi_2(\vec{x}) \land \phi_3(\vec{x})}{\vec{a}(\vec{x})}\\
    \leadsto_e & \prob{\vec{x}' = \vec{a}^n(\vec{x}) \land \phi_1(\vec{x})}{\phi_1(\vec{x})}{\phi_2(\vec{x}) \land \phi_3(\vec{x})}{\vec{a}(\vec{x})}.
  \end{align*}
  Next, if $\phi_2 \neq \true$, then \Cref{thm:conditional-one-way} applies to $\langle \phi_2, \vec{a} \rangle$ with $\check\phi \Def \phi_1$ due to \eqref{eq:simulate-three-way2} and we obtain
  \begin{align*}
    &\prob{\vec{x}' = \vec{a}^n(\vec{x}) \land \phi_1(\vec{x})}{\phi_1(\vec{x})}{\phi_2(\vec{x}) \land \phi_3(\vec{x})}{\vec{a}(\vec{x})}\\
    \leadsto_e& \prob{\vec{x}' = \vec{a}^n(\vec{x}) \land \phi_1(\vec{x}) \land \phi_2(\vec{a}^{n-1}(\vec{x}))}{\phi_1(\vec{x}) \land \phi_2(\vec{x})}{\phi_3(\vec{x})}{\vec{a}(\vec{x})}.
  \end{align*}
  Finally, if $\phi_3 \neq \true$, then \Cref{thm:conditional-recurrent} applies to $\langle \phi_3, \vec{a} \rangle$ with $\check\phi \Def \phi_1 \land \phi_2$ due to \eqref{eq:simulate-three-way3} and we obtain
  \begin{align*}
    &\prob{\vec{x}' = \vec{a}^n(\vec{x}) \land \phi_1(\vec{x}) \land \phi_2(\vec{a}^{n-1}(\vec{x}))}{\phi_1(\vec{x}) \land \phi_2(\vec{x})}{\phi_3(\vec{x})}{\vec{a}(\vec{x})}\\
    \leadsto_e &\prob{\vec{x}' = \vec{a}^n(\vec{x}) \land \phi_1(\vec{x}) \land \phi_2(\vec{a}^{n-1}(\vec{x})) \land \phi_3(\vec{x})}{\phi(\vec{x})}{\true}{\vec{a}(\vec{x})} \\
    =& \prob{\psi(\vec{x})}{\phi(\vec{x})}{\true}{\vec{a}(\vec{x})}.
  \end{align*}
  \qed
}

Thus, there is no need for a conditional variant of \emph{acceleration via monotonicity}.
Note that combining \Cref{thm:conditional-one-way,thm:conditional-recurrent} with our calculus is also useful for loops where \emph{acceleration via monotonicity} is inapplicable.

\begin{example}
  Consider the following loop, which can be accelerated by splitting its guard into one invariant and two converse invariants.
  \begin{equation}
    \label{eq:beyond-three-way}\tag{\ensuremath{\TT_{2\text{-}c\text{-}invs}}}
    \mWhile{x_1 > 0 \land x_2 > 0 \land x_3 > 0}{\mat{x_1\\x_2\\x_3} \assign \mat{x_1 - 1\\x_2+x_1\\x_3-x_2}}
  \end{equation}
  Let
  \begin{align*}
    \phi_{2\text{-}c\text{-}invs} &\Def x_1 > 0 \land x_2 > 0 \land x_3 > 0,\\
    \vec{a}_{2\text{-}c\text{-}invs} &\Def \mat{x_1 - 1\\x_2+x_1\\x_3-x_2},\\
    \psi^{init}_{2\text{-}c\text{-}invs} &\Def \vec{x}' = \vec{a}_{2\text{-}c\text{-}invs}^n(\vec{x}),
  \end{align*}
  and let $x_i^{(m)}$ be the $i^{th}$ component of $\vec{a}_{2\text{-}c\text{-}invs}^{m}(\vec{x})$.
  Starting with the canonical acceleration problem of \ref{eq:beyond-three-way}, we obtain:
  \begin{align*}
    & \prob{\psi^{init}_{2\text{-}c\text{-}invs}}{\true}{\phi_{2\text{-}c\text{-}invs}}{\vec{a}_{2\text{-}c\text{-}invs}} \\
    \overset{\Cref{thm:conditional-one-way}}{\leadsto_e} & \prob{\psi^{init}_{2\text{-}c\text{-}invs} \land x_1^{(n-1)} > 0}{x_1 > 0}{x_2 > 0 \land x_3 > 0}{\vec{a}_{2\text{-}c\text{-}invs}} \\
    \overset{\Cref{thm:conditional-recurrent}}{\leadsto_e} & \prob{\psi^{init}_{2\text{-}c\text{-}invs} \land x_1^{(n-1)} > 0 \land x_2 > 0}{x_1 > 0 \land x_2 > 0}{x_3 > 0}{\vec{a}_{2\text{-}c\text{-}invs}} \\
    \overset{\Cref{thm:conditional-one-way}}{\leadsto_e} & \prob{\psi^{init}_{2\text{-}c\text{-}invs} \land x_1^{(n-1)} \! > 0 \land x_2 > 0 \land x_3^{(n-1)} \! > 0}{\phi_{2\text{-}c\text{-}invs}}{\true}{\vec{a}_{2\text{-}c\text{-}invs}}
  \end{align*}
\end{example}

Finally, we present a variant of \Cref{thm:meter} for conditional acceleration.
The idea is similar to the approach for deducing metering functions of the form $\vec{x} \mapsto \charfun{\check{\phi}}(\vec{x}) \cdot f(\vec{x})$ from \cite{journal} (see \Cref{sec:metering} for details).
But in contrast to \cite{journal}, in our setting the ``conditional'' part $\check{\phi}$ does not need to be an invariant of the loop.
\begin{theorem}[Conditional Acceleration via Metering Functions]
  \label{thm:conditional-metering}
  Let $\mf: \ZZ^d \to \QQ$.
  If
  \begin{align}
    \report{\label{eq:conditional-metering-decrease}}
    \submission{\notag}
    \check{\phi}(\vec{x}) \land \phantom{\neg}\chi(\vec{x}) & \implies \mf(\vec{x}) - \mf(\vec{a}(\vec{x})) \leq 1 & \text{and} \\
    \report{\label{eq:conditional-metering-bound}}
    \submission{\notag}
    \check{\phi}(\vec{x}) \land \neg\chi(\vec{x}) & \implies \mf(\vec{x}) \leq 0,
  \end{align}
  then the following conditional acceleration technique is sound:
  \[
    (\langle \chi, \vec{a} \rangle, \check{\phi}) \mapsto \vec{x}' = \vec{a}^n(\vec{x}) \land \chi(\vec{x}) \land n < \mf(\vec{x}) + 1
  \]
\end{theorem}
\makeproof{thm:conditional-metering}{
  We need to prove
  \begin{equation}
    \label{eq:conditional-metering-ih}
    \vec{x} \longrightarrow^m_{\langle \check\phi, \vec{a} \rangle} \vec{a}^{m}(\vec{x}) \land \chi(\vec{x}) \land m < \mf(\vec{x}) + 1 \implies \vec{x} \longrightarrow^m_{\langle \chi, \vec{a} \rangle} \vec{a}^{m}(\vec{x})
  \end{equation}
  for all $m > 0$.
  We use induction on $m$.
  If $m=1$, then
  \begin{align*}
    & \vec{x} \longrightarrow^m_{\langle \check\phi, \vec{a} \rangle} \vec{a}^{m}(\vec{x}) \land \chi(\vec{x}) \land m < \mf(\vec{x}) + 1 \\
    \implies & \vec{x} \longrightarrow_{\langle \chi, \vec{a} \rangle} \vec{a}(\vec{x})\\
    \iff & \vec{x} \longrightarrow^m_{\langle \chi, \vec{a} \rangle} \vec{a}^m(\vec{x}) \tag{as $m=1$}
  \end{align*}
  In the induction step, assume
  \begin{equation}
    \label{eq:conditional-metering-is}
    \vec{x} \longrightarrow^{m+1}_{\langle \check{\phi}, \vec{a} \rangle} \vec{a}^{m+1}(\vec{x}) \land \chi(\vec{x}) \land m < \mf(\vec{x}).
  \end{equation}
  Then we have:
  \begin{align*}
    & \eqref{eq:conditional-metering-is} \\
    \implies & \vec{x} \longrightarrow^{m}_{\langle \check{\phi}, \vec{a} \rangle} \vec{a}^{m}(\vec{x}) \land \chi(\vec{x}) \land m < \mf(\vec{x}) \\
    \implies & \vec{x} \longrightarrow^m_{\langle \check{\phi}, \vec{a} \rangle} \vec{a}^{m}(\vec{x}) \land m < \mf(\vec{x}) \land \vec{x} \longrightarrow^m_{\langle \chi, \vec{a} \rangle} \vec{a}^{m}(\vec{x}) \tag{due to the induction hypothesis \eqref{eq:conditional-metering-ih}} \\
    \implies & m < \mf(\vec{x}) \land \vec{x} \longrightarrow^m_{\langle \chi, \vec{a} \rangle} \vec{a}^{m}(\vec{x})  \land \forall i \in [0,m-1].\ \left( \check{\phi}(\vec{a}^i(\vec{x})) \land \chi(\vec{a}^i(\vec{x})) \right)\\
    \implies & m < \mf(\vec{x}) \land \vec{x} \longrightarrow^m_{\langle \chi, \vec{a} \rangle} \vec{a}^{m}(\vec{x}) \land \mf(\vec{x}) - \mf(\vec{a}^m(\vec{x})) \leq m \tag{due to \eqref{eq:conditional-metering-decrease}}\\
    \implies & \vec{x} \longrightarrow^m_{\langle \chi, \vec{a} \rangle} \vec{a}^{m}(\vec{x}) \land \mf(\vec{a}^m(\vec{x})) > 0 \tag{as $m < \mf(\vec{x})$, cf.\ \eqref{eq:conditional-metering-is}}
  \end{align*}
  Note that \eqref{eq:conditional-metering-bound} is equivalent to
  \[
    \mf(\vec{x}) > 0 \implies \neg\check{\phi}(\vec{x}) \lor \chi(\vec{x}).
  \]
  Thus, from $\mf(\vec{a}^m(\vec{x})) > 0$ we get $\neg\check{\phi}(\vec{a}^m(\vec{x})) \lor \chi(\vec{a}^m(\vec{x}))$.
  Moreover, as \eqref{eq:conditional-metering-is} implies $\check{\phi}(\vec{a}^m(\vec{x}))$, we obtain $\chi(\vec{a}^m(\vec{x}))$.
  Together with $\vec{x} \longrightarrow^m_{\langle \chi, \vec{a} \rangle} \vec{a}^{m}(\vec{x})$, this implies $\vec{x} \longrightarrow^{m+1}_{\langle \chi, \vec{a} \rangle} \vec{a}^{m+1}(\vec{x})$, as desired.
  \qed
}
\section{Acceleration via Eventual Monotonicity}
\label{sec:accel}

The combination of the calculus from \Cref{sec:integration} and the conditional acceleration techniques from \Cref{sec:conditional} still fails to handle certain interesting classes of loops.
Thus, to improve the applicability of our approach we now present two new acceleration techniques based on \emph{eventual} monotonicity.

\subsection{Acceleration via Eventual Decrease}

All (combinations of) techniques presented so far fail for the following example.
\begin{equation}
  \label{eq:phases}\tag{\ensuremath{\TT_{ev\text{-}dec}}}
  \mWhile{x_1 > 0}{\mat{x_1\\x_2} \assign \mat{x_1 + x_2\\x_2 - 1}}
\end{equation}
The reason is that $x_1$ does not behave monotonically, i.e., $x_1 > 0$ is neither an invariant nor a converse invariant.
Essentially, \ref{eq:phases} proceeds in two phases:
In the first (optional) phase, $x_2$ is positive and hence the value of $x_1$ is monotonically increasing.
In the second phase, $x_2$ is non-positive and consequently the value of $x_1$ decreases (weakly) monotonically.
The crucial observation is that once the value of $x_1$ decreases, it can never increase again.
Thus, despite the non-monotonic behavior of $x_1$, it suffices to require that $x_1 > 0$ holds before the first and before the $n^{th}$ loop iteration to ensure that the loop can be iterated at least $n$ times.
\begin{theorem}[Acceleration via Eventual Decrease]
  \label{thm:ev-dec}
  If $\phi(\vec{x}) \equiv \bigwedge_{i=1}^k C_i$ where each $C_i$ contains an inequation $\expr_i(\vec{x}) > 0$ such that
  \[
    \expr_i(\vec{x}) \geq \expr_i(\vec{a}(\vec{x})) \implies \expr_i(\vec{a}(\vec{x})) \geq \expr_i(\vec{a}^2(\vec{x})),
  \]
  then the following acceleration technique is sound:
  \[
    \ref{loop} \mapsto \vec{x}' = \vec{a}^n(\vec{x}) \land \bigwedge_{i=1}^k \left(\expr_i(\vec{x}) > 0 \land \expr_i(\vec{a}^{n-1}(\vec{x})) > 0\right)
  \]
  If $C_i \equiv \expr_i > 0$ for all $i \in [1,k]$, then it is exact.
\end{theorem}
\makeproof{thm:ev-dec}{We will prove the more general \Cref{thm:ev-dec-cond} later in this section. \qed}

With \Cref{thm:ev-dec}, we can accelerate \ref{eq:phases} to
\[
  \mat{x_1'\\x_2'} = \mat{\tfrac{n-n^2}{2} + x_2 \cdot n + x_1\\x_2 - n} \land x_1 > 0 \land \tfrac{n-1-(n-1)^2}{2} + x_2 \cdot (n-1) + x_1 > 0
\]
as we have
\[
  (x_1 \geq x_1 + x_2) \equiv (0 \geq x_2) \implies (0 \geq x_2 - 1) \equiv (x_1 + x_2 \geq x_1 + x_2 + x_2 - 1).
\]
Turning \Cref{thm:ev-dec} into a conditional acceleration technique is straightforward.
\begin{theorem}[Conditional Acceleration via Eventual Decrease]
  \label{thm:ev-dec-cond}
  If we have $\chi(\vec{x}) \equiv \bigwedge_{i=1}^k C_i$ where each $C_i$ contains an inequation $\expr_i(\vec{x}) > 0$ such that
  \begin{equation}
    \label{eq:ev-dec-cond-pre}
    \check\phi(\vec{x}) \land \expr_i(\vec{x}) \geq \expr_i(\vec{a}(\vec{x})) \implies \expr_i(\vec{a}(\vec{x})) \geq \expr_i(\vec{a}^2(\vec{x})),
  \end{equation}
  then the following conditional acceleration technique is sound:
  \begin{equation}
    \submission{\notag}
    \label{eq:ev-dec-cond}
    (\langle \chi,\vec{a} \rangle, \check\phi) \mapsto \vec{x}' = \vec{a}^n(\vec{x}) \land \bigwedge_{i=1}^k \left( \expr_i(\vec{x}) > 0 \land \expr_i(\vec{a}^{n-1}(\vec{x})) > 0 \right)
  \end{equation}
  If $C_i \equiv \expr_i > 0$ for all $i \in [1,k]$, then it is exact.
\end{theorem}
\makeproof{thm:ev-dec-cond}{
  For soundness, we need to show
  \begin{multline}
    \label{eq:ev-dec-goal-orig}
    \vec{x} \longrightarrow^n_{\langle \check\phi,\vec{a} \rangle} \vec{a}^n(\vec{x}) \land \bigwedge_{i=1}^k \left( \expr_i(\vec{x}) > 0 \land \expr_i(\vec{a}^{n-1}(\vec{x})) > 0 \right)\\
    \implies \vec{x} \longrightarrow^n_{\langle \chi, \vec{a}\rangle} \vec{a}^n(\vec{x}).
  \end{multline}
  Assume
  \begin{equation}
    \label{eq:ev-dec-assumption-1}
    \vec{x} \longrightarrow^n_{\langle \check\phi,\vec{a} \rangle} \vec{a}^n(\vec{x}) \land \bigwedge_{i=1}^k \left( \expr_i(\vec{x}) > 0 \land \expr_i(\vec{a}^{n-1}(\vec{x})) > 0 \right).
  \end{equation}
  This implies
  \begin{equation}
    \label{eq:ev-dec-assumption2}
    \bigwedge_{i=0}^{n-1} \check\phi(\vec{a}^i(\vec{x})).
  \end{equation}
  In the following, we show
  \begin{equation}
    \label{eq:ev-dec-goal}
    \bigwedge_{i=1}^k\bigwedge_{m=0}^{n-1} \expr_i(\vec{a}^m(\vec{x})) \geq \min(\expr_i(\vec{x}), \expr_i(\vec{a}^{n-1}(\vec{x}))).
  \end{equation}
  Then the claim follows, as we have
  \begin{align*}
    &\bigwedge_{i=1}^k\bigwedge_{m=0}^{n-1} \expr_i(\vec{a}^m(\vec{x})) \geq \min(\expr_i(\vec{x}), \expr_i(\vec{a}^{n-1}(\vec{x})))\\
    \implies& \bigwedge_{i=1}^k\bigwedge_{m=0}^{n-1} \expr_i(\vec{a}^m(\vec{x})) \geq 0 \tag{due to \eqref{eq:ev-dec-assumption-1}}\\
    \implies& \bigwedge_{m=0}^{n-1} \chi(\vec{a}^m(\vec{x})) \tag{by definition of $\expr_i$}\\
    \implies& \vec{x} \longrightarrow^n_{\langle \chi, \vec{a}\rangle} \vec{a}^n(\vec{x}).
  \end{align*}
  Let $i$ be arbitrary but fixed, let $\expr = \expr_i$, and let $j$ be the minimal natural number with
  \begin{equation}
    \label{eq:ev-dec-assumption}
    \expr(\vec{a}^j(\vec{x})) = \max\{\expr(\vec{a}^m(\vec{x})) \mid m \in [0,n-1]\}.
  \end{equation}
  We first prove
  \begin{equation}
    \label{eq:ev-dec-ih}
    \expr(\vec{a}^{m}(\vec{x})) < \expr(\vec{a}^{m+1}(\vec{x}))
  \end{equation}
  for all $m \in [0,j-1]$ by backward induction on $m$.
  If $m=j-1$, then
  \begin{align*}
    & \expr(\vec{a}^{m}(\vec{x})) \\
    = {}& \expr(\vec{a}^{j-1}(\vec{x})) \tag{as $m=j-1$} \\
    < {}& \expr(\vec{a}^{j}(\vec{x})) \tag{due to \eqref{eq:ev-dec-assumption} as $j$ is minimal} \\
    = {}& \expr(\vec{a}^{m+1}(\vec{x})). \tag{as $m=j-1$}
  \end{align*}

  For the induction step, note that \eqref{eq:ev-dec-cond-pre} implies
  \begin{equation}
    \label{eq:ev-dec-cond-pre2}
    \expr(\vec{a}(\vec{x})) < \expr(\vec{a}^2(\vec{x})) \implies \neg\check\phi(\vec{x}) \lor \expr(\vec{x}) < \expr(\vec{a}(\vec{x})).
  \end{equation}
  In the induction step, we have
  \begin{align*}
    &\expr(\vec{a}^{m}(\vec{x})) < \expr(\vec{a}^{m+1}(\vec{x})) \tag{due to the induction hypothesis \eqref{eq:ev-dec-ih}} \\
    \implies & \neg\check\phi(\vec{a}^{m}(\vec{x})) \lor \expr(\vec{a}^{m-1}(\vec{x})) < \expr(\vec{a}^{m}(\vec{x})) \tag{by \eqref{eq:ev-dec-cond-pre2}} \\
    \implies & \expr(\vec{a}^{m-1}(\vec{x})) < \expr(\vec{a}^{m}(\vec{x})). \tag{by \eqref{eq:ev-dec-assumption2}}
  \end{align*}

  Now we prove
  \begin{equation}
    \label{eq:ev-dec-ih2}
    \expr(\vec{a}^m(\vec{x})) \geq \expr(\vec{a}^{m+1}(\vec{x}))
  \end{equation}
  for all $m \in [j,n-1]$ by induction on $m$.
  If $m=j$, then
  \begin{align*}
    & \expr(\vec{a}^{m}(\vec{x})) \\
    = {}& \expr(\vec{a}^{j}(\vec{x})) \tag{as $m=j$} \\
    = {}& \max\{\expr(\vec{a}^m(\vec{x})) \mid m \in [0,n-1]\} \tag{due to \eqref{eq:ev-dec-assumption}}\\
    \geq {}& \expr(\vec{a}^{j+1}(\vec{x})) \\
    = {}& \expr(\vec{a}^{m+1}(\vec{x})).\tag{as $m=j$}
  \end{align*}

  In the induction step, we have
  \begin{align*}
    &\expr(\vec{a}^{m}(\vec{x})) \geq \expr(\vec{a}^{m+1}(\vec{x})) \tag{due to the induction hypothesis \eqref{eq:ev-dec-ih2}} \\
    \implies & \expr(\vec{a}^{m+1}(\vec{x})) \geq \expr(\vec{a}^{m+2}(\vec{x})) \tag{due to \eqref{eq:ev-dec-assumption2} and \eqref{eq:ev-dec-cond-pre}}.
  \end{align*}

  As \eqref{eq:ev-dec-ih} and \eqref{eq:ev-dec-ih2} imply
  \[
    \bigwedge_{m=0}^{n-1}\expr(\vec{a}^m(\vec{x})) \geq \min(\expr(\vec{x}), \expr(\vec{a}^{n-1}(\vec{x}))),
  \]
  this finishes the proof of \eqref{eq:ev-dec-goal} and hence shows \eqref{eq:ev-dec-goal-orig}.

  For exactness, assume $\chi(\vec{x}) \Def \bigwedge_{i=1}^k \expr_i(\vec{x}) > 0$.
  We have
  \begin{align*}
    &\vec{x} \longrightarrow^n_{\langle \chi \land \check\phi, \vec{a}\rangle} \vec{a}^n(\vec{x})\\
    \implies& \chi(\vec{x}) \land \chi(\vec{a}^{n-1}(\vec{x})) \\
    \iff& \bigwedge_{i=1}^k \left(\expr_i(\vec{x}) > 0 \land \expr_i(\vec{a}^{n-1}(\vec{x})) > 0\right).
  \end{align*}
  \qed
}

\begin{example}
  Consider the following variant of \ref{eq:phases}.
  \[
    \mWhile{x_1 > 0 \land x_3 > 0}{\mat{x_1\\x_2\\x_3} \assign \mat{x_1 + x_2\\x_2 - x_3\\x_3}}
  \]
  Starting with its canonical acceleration problem, we get
  \[
    \begin{array}{cl}
      & \prob{\vec{x}' = \vec{a}^n(\vec{x})}{\top}{x_1 > 0 \land x_3 > 0}{\vec{a} \Def \mat{x_1 + x_2\\x_2 - x_3\\x_3}}\\[.5em]
      \overset{\Cref{thm:conditional-recurrent}}{\leadsto_e} & \hspace{1.25pt} \prob{\vec{x}' = \vec{a}^n(\vec{x}) \land x_3 > 0}{x_3 > 0}{x_1 > 0}{\vec{a}} \\[.5em]
      \overset{\Cref{thm:ev-dec-cond}}{\leadsto_e} & \prob{\vec{x}' = \vec{a}^n(\vec{x}) \land x_3 > 0 \land x_1 > 0 \land x_1^{(n-1)} > 0}{x_3 > 0 \land x_1 > 0}{\top}{\vec{a}}
    \end{array}
  \]
  where the second step can be performed via \Cref{thm:ev-dec-cond} as
  \[
    (\check\phi(\vec{x}) \land \expr(\vec{x}) \geq \expr(\vec{a}(\vec{x}))) \equiv (x_3 > 0 \land x_1 \geq x_1 + x_2) \equiv (x_3 > 0 \land 0 \geq x_2)\\
  \]
  implies
  \[
    (0 \geq x_2 - x_3) \equiv (x_1 + x_2 \geq x_1 + x_2 + x_2 - x_3) \equiv (\expr(\vec{a}(\vec{x})) \geq \expr(\vec{a}^2(\vec{x}))).
  \]
\end{example}

\subsection{Acceleration via Eventual Increase}

Still, all (combinations of) techniques presented so far fail for
\begin{equation}
  \label{eq:phases2}\tag{\ensuremath{\TT_{ev\text{-}inc}}}
  \mWhile{x_1 > 0}{\mat{x_1\\x_2} \assign \mat{x_1 + x_2\\x_2 + 1}}.
\end{equation}
As in the case of \ref{eq:phases}, the value of $x_1$ does not behave monotonically, i.e., $x_1 > 0$ is neither an invariant nor a converse invariant.
However, this time $x_1$ is eventually \emph{increasing}, i.e., once $x_1$ starts to grow, it never decreases again.
Thus, in this case it suffices to require that $x_1$ is positive and (weakly) increasing.

\begin{theorem}[Acceleration via Eventual Increase]
  \label{thm:ev-inc}
  If $\phi(\vec{x}) \equiv \bigwedge_{i=1}^k C_i$ where each $C_i$ contains an inequation $\expr_i(\vec{x}) > 0$ such that
  \[
    \expr_i(\vec{x}) \leq \expr_i(\vec{a}(\vec{x})) \implies \expr_i(\vec{a}(\vec{x})) \leq \expr_i(\vec{a}^2(\vec{x})),
  \]
  then the following acceleration technique is sound:
  \[
    \ref{loop} \mapsto \vec{x}' = \vec{a}^n(\vec{x}) \land \bigwedge_{i=1}^k 0 < \expr_i(\vec{x}) \leq \expr_i(\vec{a}(\vec{x}))
  \]
\end{theorem}
\makeproof{thm:ev-inc}{We prove the more general \Cref{thm:conditional-ev-inc} later in this section. \qed}

With \Cref{thm:ev-inc}, we can accelerate \ref{eq:phases2} to
\begin{equation}
  \label{psi:phases2}\tag{\ensuremath{\psi_{ev\text{-}inc}}}
  \mat{x_1'\\x_2'} = \mat{\tfrac{n^2-n}{2} + x_2 \cdot n + x_1 \\ x_2 + n} \land 0 < x_1 \leq x_1 + x_2
\end{equation}
as we have
\[
  (x_1 \leq x_1 + x_2) \equiv (0 \leq x_2) \implies (0 \leq x_2 + 1) \equiv (x_1 + x_2 \leq x_1 + x_2 + x_2 + 1).
\]
However, \Cref{thm:ev-inc} is \emph{not} exact, as the resulting formula only covers program runs where each $\expr_i$ behaves monotonically.
So \ref{psi:phases2} only covers those runs of \ref{eq:phases2} where the initial value of $x_2$ is non-negative.
Again, turning \Cref{thm:ev-inc} into a conditional acceleration technique is straightforward.

\begin{theorem}[Conditional Acceleration via Eventual Increase]
  \label{thm:conditional-ev-inc}
  If we have $\chi(\vec{x}) \equiv \bigwedge_{i=1}^k C_i$ where each $C_i$ contains an inequation $\expr_i(\vec{x}) > 0$ such that
  \begin{equation}
    \label{eq:ev-inc-pre}
    \check{\phi}(\vec{x}) \land \expr_i(\vec{x}) \leq \expr_i(\vec{a}(\vec{x})) \implies \expr_i(\vec{a}(\vec{x})) \leq \expr_i(\vec{a}^2(\vec{x})),
  \end{equation}
  then the following conditional acceleration technique is sound:
  \begin{equation}
    \submission{\notag}
    \label{eq:ev-inc}
    (\langle \chi, \vec{a} \rangle, \check{\phi}) \mapsto \vec{x}' = \vec{a}^n(\vec{x}) \land \bigwedge_{i=1}^k 0 < \expr_i(\vec{x}) \leq \expr_i(\vec{a}(\vec{x}))
  \end{equation}
\end{theorem}
\makeproof{thm:conditional-ev-inc}{
  We need to show
  \[
    \vec{x} \longrightarrow_{\langle \check\phi, \vec{a} \rangle}^n \vec{a}^n(\vec{x}) \land \bigwedge_{i=1}^k 0 < \expr_i(\vec{x}) \leq \expr_i(\vec{a}(\vec{x})) \implies \vec{x} \longrightarrow^n_{\langle \chi, \vec{a} \rangle} \vec{a}^n(\vec{x}).
  \]
  Due to $\vec{x} \longrightarrow_{\langle \check\phi, \vec{a} \rangle}^n \vec{a}^n(\vec{x})$, we have
  \begin{equation}
    \label{eq:ev-inc-check}
    \bigwedge_{j=0}^{n-1} \check\phi(\vec{a}^j(\vec{x})).
  \end{equation}
  Let $i$ be arbitrary but fixed and assume
  \begin{equation}
    \label{eq:ev-inc-assumption}
    0 < \expr_i(\vec{x}) \leq \expr_i(\vec{a}(\vec{x})).
  \end{equation}
  We prove
  \begin{equation}
    \label{eq:ev-inc-ih}
    \expr_i(\vec{a}^m(\vec{x})) \leq \expr_i(\vec{a}^{m+1}(\vec{x}))
  \end{equation}
  for all $m < n $ by induction on $m$.
  Then we get
  \[
    0 < \expr_i(\vec{a}^m(\vec{x}))
  \]
  and thus $\phi(\vec{a}^m(\vec{x}))$ for all $m < n$ due to \eqref{eq:ev-inc-assumption} and hence the claim follows.
  If $m=0$, then
  \[
    \expr^m_i(\vec{x}) = \expr_i(\vec{x}) \leq \expr_i(\vec{a}(\vec{x})) = \expr_i(\vec{a}^{m+1}(\vec{x})). \tag{due to \eqref{eq:ev-inc-assumption}}
  \]
  In the induction step, note that \eqref{eq:ev-inc-check} implies
  \[
    \check\phi(\vec{a}^m(\vec{x}))
  \]
  as $m<n$.
  Together with the induction hypothesis \eqref{eq:ev-inc-ih}, we get
  \[
    \check\phi(\vec{a}^m(\vec{x})) \land \expr_i(\vec{a}^m(\vec{x})) \leq \expr_i(\vec{a}^{m+1}(\vec{x})).
  \]
  By \eqref{eq:ev-inc-pre}, this implies
  \[
    \expr_i(\vec{a}^{m+1}(\vec{x})) \leq \expr_i(\vec{a}^{m+2}(\vec{x})),
  \]
  as desired. \qed
}

\begin{example}
  Consider the following variant of \ref{eq:phases2}.
  \[
    \mWhile{x_1 > 0 \land x_3 > 0}{\mat{x_1\\x_2\\x_3} \assign \mat{x_1 + x_2\\x_2 + x_3\\x_3}}
  \]
  Starting with its canonical acceleration problem, we get
  \[
    \begin{array}{cl}
      & \prob{\vec{x}' = \vec{a}^n(\vec{x})}{\top}{x_1 > 0 \land x_3 > 0}{\vec{a} \Def \mat{x_1 + x_2\\x_2 + x_3\\x_3}}\\[.5em]
      \overset{\Cref{thm:conditional-recurrent}}{\leadsto_e} & \hspace{1.25pt} \prob{\vec{x}' = \vec{a}^n(\vec{x}) \land x_3 > 0}{x_3 > 0}{x_1 > 0}{\vec{a}} \\[.5em]
      \overset{\Cref{thm:conditional-ev-inc}}{\leadsto} & \hspace{1.25pt} \prob{\vec{x}' = \vec{a}^n(\vec{x}) \land x_3 > 0 \land 0 < x_1 \leq x_1 + x_2}{x_3 > 0 \land x_1 > 0}{\top}{\vec{a}}
    \end{array}
  \]
  where the second step can be performed via \Cref{thm:conditional-ev-inc} as
  \[
    (\check\phi(\vec{x}) \land \expr(\vec{x}) \leq \expr(\vec{a}(\vec{x}))) \equiv (x_3 > 0 \land x_1 \leq x_1 + x_2) \equiv (x_3 > 0 \land 0 \leq x_2)\\
  \]
  implies
  \[
    (0 \leq x_2 + x_3) \equiv (x_1 + x_2 \leq x_1 + x_2 + x_2 + x_3) \equiv (\expr(\vec{a}(\vec{x})) \leq \expr(\vec{a}^2(\vec{x}))).
  \]
\end{example}

We also considered versions of \Cref{thm:ev-dec-cond,thm:conditional-ev-inc} where the inequations in \eqref{eq:ev-dec-cond-pre} resp.\ \eqref{eq:ev-inc-pre} are strict, but this did not lead to an improvement in our experiments.
Moreover, we experimented with a variant of \Cref{thm:conditional-ev-inc} that splits the loop under consideration into two consecutive loops, accelerates them independently, and composes the results.
While such an approach can accelerate loops like \ref{psi:phases2} exactly, the impact on our experimental results was minimal.
Thus, we postpone an in-depth investigation of this idea to future work.
\section{Related Work}
\label{sec:related}

Acceleration-like techniques are also used in \emph{over-ap\-prox\-i\-mat\-ing} settings (see, e.g., \cite{kincaid15,gonnord06,jeannet14,madhukar15,strejcek12,silverman19,kincaid17,schrammel14}), whereas we consider \emph{exact} and \emph{under-ap\-prox\-i\-mat\-ing} loop acceleration techniques.
As many related approaches have already been discussed in \Cref{sec:monotonic}, we only mention two more techniques here.

First, \cite{finite-monid,bozga10} presents an exact acceleration technique for \emph{finite monoid affine transformations} (FMATs), i.e., loops with linear arithmetic whose body is of the form $\vec{x} \assign A\vec{x} + \vec{b}$ where $\{A^i \mid i \in \NN\}$ is finite.
For such loops, Pres\-burger-Arithmetic is sufficient to construct an equivalent formula $\psi$, i.e., it can be expressed in a decidable logic.
In general, this is clearly not the case for the techniques presented in the current paper (which may even synthesize non-polynomial closed forms, see \ref{loop:exp}).
As a consequence and in contrast to our technique, this approach cannot handle loops where the values of variables grow super-linearly (i.e., it cannot handle examples like \ref{eq:three-way-ex}).
Implementations are available in the tools \tool{FAST} \cite{fast} and \tool{Flata} \cite{hojjat12}.
Further theoretical results on linear transformations whose $n$-fold closure is definable in (extensions of) Presburger-Arithmetic can be found in \cite{Boigelot03}.

% In future work, integrating the technique from \cite{finite-monid,bozga10} into our framework would allow for simplifying acceleration problems $\prob{\psi}{\check\phi}{\hat\phi_\ell \land \hat\phi_{n\ell}}{\vec{a}}$ where $\langle \hat\phi_\ell, \vec{a} \rangle$ is an FMAT, but $\hat\phi_{n\ell}$ contains non-linear inequations.
% %
% Then the resulting acceleration problem $\prob{\psi \land \ldots}{\check\phi \land \hat\phi_\ell}{\hat\phi_{n\ell}}{\vec{a}}$ could be further processed by other techniques that support non-linear arithmetic.

Second, \cite{bozga09a} shows that \emph{octagonal relations} can be accelerated exactly and in \cite{octagonsP}, it is proven that such relations can even be accelerated in polynomial time.
This generalizes earlier results for \emph{difference bound constraints} \cite{differenceBounds}.
As in the case of FMATs, the resulting formula can be expressed in Presburger-Arithmetic.
Octagonal relations are defined by a finite conjunction $\xi$ of inequations of the form $\pm x \pm y \leq c$, $x,y \in \vec{x} \cup \vec{x}'$, $c \in \ZZ$.
Then $\xi$ induces the relation $\vec{x} \longrightarrow_\xi \vec{x}' \iff \xi(\vec{x}, \vec{x}')$.
So in contrast to the loops considered in the current paper where $\vec{x}'$ is uniquely determined by $\vec{x}$, octagonal relations can represent non-deterministic programs.
Therefore and due to the restricted form of octagonal relations, the work from \cite{bozga09a,octagonsP} is orthogonal to ours.

\section{Implementation and Experiments}
\label{sec:experiments}

We prototypically implemented our approach in our open-source \underline{Lo}op \underline{A}c\-cel\-er\-a\-tion \underline{T}ool \loat \cite{ijcar16,fmcad19,journal}:
\begin{center}
  \url{https://github.com/aprove-developers/LoAT/tree/tacas20}
\end{center}
It uses \tool{Z3} \cite{z3} to check implications and \tool{PURRS} \cite{purrs} to compute closed forms.

For technical reasons, the closed forms computed by \loat are valid only if $n>0$, whereas \Cref{def:closed} requires them to be valid for all $n \in \NN$.
The reason is that \tool{PURRS} has only limited support for initial conditions.
In the future, we plan to use a different recurrence solver to circumvent this problem.
Thus, \loat's results are only correct for all $n>1$ (instead of all $n > 0$).
Moreover, \loat can currently compute closed forms only if the loop body is \emph{triangular}, meaning that each $a_i$ is an expression over $x_1,\ldots,x_i$.
The reason is that \tool{PURRS} cannot solve \emph{systems} of recurrence equations, but only a single recurrence equation at a time.
However, \loat failed to compute closed forms for just $26$ out of $1511$ loops in our experiments, i.e., this appears to be a minor restriction in practice.
Furthermore, \emph{conditional acceleration via metering functions} has not yet been integrated into the implementation of our calculus.
While \loat can synthesize formulas with non-polynomial arithmetic, it cannot yet parse them, i.e., the input is restricted to polynomials.
Finally, \loat does not yet support disjunctive loop conditions.

Apart from these differences, our implementation closely follows the current paper.
It repeatedly applies the conditional acceleration techniques from \Cref{sec:conditional,sec:accel} with the following priorities: $\Cref{thm:conditional-recurrent} > \Cref{thm:conditional-one-way} > \Cref{thm:ev-dec-cond} > \Cref{thm:conditional-ev-inc}$.

To evaluate our approach, we extracted $1511$ loops with conjunctive guards from the category \emph{Termination of Integer Transition Systems} of the \emph{Termination Problems Database} \cite{tpdb}, the benchmark collection which is used at the annual \emph{Termination and Complexity Competition} \cite{termcomp}, as follows:
\begin{enumerate}
\item We parsed all examples with \loat and extracted each single-path loop with conjunctive guard (resulting in $3829$ benchmarks).
\item We removed duplicates by checking syntactic equality (resulting in $2825$ benchmarks).
\item We removed loops whose runtime is trivially constant using an incomplete check (resulting in $1733$ benchmarks).
\item We removed loops which do not admit any terminating runs, i.e., loops where \Cref{thm:recurrent} applies (resulting in $1511$ benchmarks).
\end{enumerate}
We compared our implementation with \loat's implementation of \emph{acceleration via monotonicity} (\Cref{thm:three-way}, \cite{fmcad19}) and its implementation of \emph{acceleration via metering functions} (\Cref{thm:meter}, \cite{ijcar16}), which also incorporates the improvements proposed in \cite{journal}.
We did not include the techniques from \Cref{thm:one-way,thm:recurrent} in our evaluation, as they are subsumed by \emph{acceleration via monotonicity}.
Furthermore, we compared with \tool{Flata} \cite{hojjat12}, which implements the techniques to accelerate FMATs and octagonal relations discussed in \Cref{sec:related}.
Note that our benchmark collection contains $16$ loops with non-linear arithmetic where \tool{Flata} is bound to fail, since it only supports linear arithmetic.
We did not compare with \tool{FAST} \cite{fast}, which uses a similar approach as the more recent tool \tool{Flata}.

All tests have been run on \tool{StarExec} \cite{starexec}.
The results can be seen in \Cref{tab1}.
They show that our novel calculus was superior to the competing techniques in our experiments.
In all but $7$ cases where our calculus successfully accelerated the given loop, the resulting formula was polynomial.
Thus, integrating our approach into existing acceleration-based verification techniques should not present major obstacles w.r.t.\ automation.

\begin{table}[ht]
\begin{minipage}{0.45\textwidth}
  \begin{center}
    \begin{tabular}{c||c|c|c|c}
             & \loat    & Monot. & Meter          & \tool{Flata} \\ \hline \hline
      exact  & 1444     & 845    & 0\footnotemark & 1231         \\ \hline
      approx & 38       & 0      & 733            & 0            \\ \hline
      fail   & 29       & 666    & 778            & 280          \\ \hline
      avg rt & 0.16s    & 0.11s  & 0.09s          & 0.47s        \\ \hline
    \end{tabular}
    \caption{}
    \label{tab1}
  \end{center}
\end{minipage}
\begin{minipage}{0.45\textwidth}
  \begin{center}
    \begin{tabular}{c||c|c|c}
             & \sout{Ev-Inc} & \sout{Ev-Dec} & \sout{Ev-Mon} \\ \hline \hline
      exact  & 1444          & 845           & 845           \\ \hline
      approx & 0             & 493           & 0             \\ \hline
      fail   & 67            & 173           & 666           \\ \hline
      avg rt & 0.15s         & 0.14s         & 0.09s         \\ \hline
    \end{tabular}
    \caption{}
    \label{tab2}
  \end{center}
\end{minipage}

\smallskip
\hfill
\begin{minipage}{0.9\textwidth}
  \begin{tabular}{ll}
    \loat:& Acceleration calculus + \Cref{thm:conditional-one-way,thm:conditional-recurrent,thm:ev-dec-cond,thm:conditional-ev-inc}\\
    Monot.:& \emph{Acceleration via Monotonicity}, \Cref{thm:three-way}\\
    Meter:& \emph{Acceleration via Metering Functions}, \Cref{thm:meter}\\
    \tool{Flata}:& The tool \tool{Flata}, see \url{http://nts.imag.fr/index.php/Flata}\\
    \sout{Ev-Inc}:& Acceleration calculus + \Cref{thm:conditional-one-way,thm:conditional-recurrent,thm:ev-dec-cond}\\
    \sout{Ev-Dec}:& Acceleration calculus + \Cref{thm:conditional-one-way,thm:conditional-recurrent,thm:conditional-ev-inc}\\
    \sout{Ev-Mon}:& Acceleration calculus + \Cref{thm:conditional-one-way,thm:conditional-recurrent}\\
    exact:& Number of examples that were accelerated \emph{exactly}\\
    approx:& Number of examples that were accelerated \emph{approximately}\\
    fail:& Number of examples that could not be accelerated\\
    avg rt:& Average runtime per example
  \end{tabular}
\end{minipage}
\hfill
\end{table}
\footnotetext{While acceleration via metering functions may be exact in some cases (see the discussion after \Cref{thm:meter}), our implementation cannot check whether this is the case.}

Furthermore, we evaluated the impact of our new acceleration techniques from \Cref{sec:accel} independently.
To this end, we once disabled \emph{acceleration via eventual increase}, \emph{acceleration via eventual decrease}, and both of them.
The results can be seen in \Cref{tab2}.
They show that our calculus does not improve over \emph{acceleration via monotonicity} if both \emph{acceleration via eventual increase} and \emph{acceleration via eventual decrease} are disabled (i.e., our benchmark collection does not contain examples like \ref{eq:beyond-three-way}).
However, enabling either \emph{acceleration via eventual decrease} or \emph{acceleration via eventual increase} resulted in a significant improvement.
Interestingly, there are many examples that can be accelerated with either of these two techniques:
When both of them were enabled, \loat (exactly or approximately) accelerated $1482$ loops.
When one of them was enabled, it accelerated $1444$ resp.\ $1338$ loops.
But when none of them was enabled, it only accelerated $845$ loops.
We believe that this is due to examples like
\[
  \mWhile{x_1 > 0 \land \ldots}{\mat{x_1\\x_2\\\ldots} \assign \mat{x_2\\x_2\\\ldots}}
\]
where \Cref{thm:ev-dec-cond} \emph{and} \Cref{thm:conditional-ev-inc} are applicable (since $x_1 \leq x_2$ implies $x_2 \leq x_2$ and $x_1 \geq x_2$ implies $x_2 \geq x_2$).

\tool{Flata} exactly accelerated 49 loops where \loat failed or approximated and \loat exactly accelerated 262 loops where \tool{Flata} failed.
So there were only 18 loops where both \tool{Flata} and the full version of our calculus failed to compute an exact result.
Among them were the only 3 examples where our implementation found a closed form, but failed anyway.
One of them was\footnote{The other two are structurally similar, but more complex.}\submission{\pagebreak}
\[
  \mWhile{x_3 > 0}{\mat{x_1\\x_2\\x_3} \assign \mat{x_1+1\\x_2-x_1\\x_3+x_2}}.
\]
Here, the updated value of $x_1$ depends on $x_1$, the update of $x_2$ depends on $x_1$ and $x_2$, and the update of $x_3$ depends on $x_2$ and $x_3$.
Hence, the closed form of $x_1$ is linear, the closed form of $x_2$ is quadratic, and the closed form of $x_3$ is cubic:
\[
  x_3^{(n)} =-\tfrac{1}{6} \cdot n^3 + \tfrac{1-x_1}{2} \cdot n^2 + \left(\tfrac{x_1}{2} + x_2 - \tfrac{1}{3}\right) \cdot n + x_3
\]
So when fixing $x_1,x_2$, and $x_3$, $x_3^{(n)}$ has up to $2$ extrema, i.e., its monotonicity may change twice.
However, our techniques based on eventual monotonicity require that the respective expressions behave monotonically once they start to de- or increase, so these techniques only allow one change of monotonicity.

This raises the question if our approach can accelerate \emph{every} loop with conjunctive guard and linear arithmetic whose closed form is a vector of (at most) quadratic polynomials with rational coefficients.
We leave this to future work.

For our benchmark collection, links to the \tool{StarExec}-jobs of our evaluation, and a pre-compiled binary (Linux, 64 bit) we refer to \cite{website}.
\section{Conclusion and Future Work}
\label{sec:conclusion}

After discussing existing acceleration techniques (\Cref{sec:monotonic}), we presented a calculus to combine acceleration techniques modularly (\Cref{sec:integration}).
Then we showed how to combine existing (\Cref{sec:conditional}) and two novel (\Cref{sec:accel}) acceleration techniques with our calculus.
This improves over prior approaches, where acceleration techniques were used independently, and may thus improve acceleration-based verification techniques \cite{underapprox15,ijcar16,fmcad19,journal,bozga09a,bozga10,iosif17} in the future.
An empirical evaluation (\Cref{sec:experiments}) shows that our approach is more powerful than state-of-the-art acceleration techniques.
Moreover, if it is able to accelerate a loop, then the result is exact (instead of just an under-approximation) in most cases.
Thus, our calculus can be used for under-approximating techniques (e.g., to find bugs or counterexamples) as well as in over-approximating settings (e.g., to prove safety or termination).

In the future, we plan to implement the missing features mentioned in \Cref{sec:experiments} and integrate our novel calculus into our own acceleration-based program analyses to prove lower bounds on the runtime complexity \cite{ijcar16,journal} and non-termination \cite{fmcad19} of integer programs.
Furthermore, our experiments indicate that integrating specialized techniques for FMATs (cf.\ \Cref{sec:related}) would improve the power of our approach, as \tool{Flata} exactly accelerated 49 loops where \tool{LoAT} failed to do so (cf.\ \Cref{sec:experiments}).
Moreover, we plan to design a \emph{loop acceleration library}, such that our technique can easily be incorporated by other verification tools.

\subsubsection*{Data Availability Statement and Acknowledgments}

The tools and datasets used for the current study are available in the Zenodo repository \cite{ae}.

I thank Carsten Fuhs, Marcel Hark, Sophie Tourret, and the anonymous reviewers for helpful feedback and comments.
Moreover, I thank Radu Iosif and Filip Konecný for their help with \tool{Flata}.

\bibliographystyle{splncs04}
\bibliography{refs,crossrefs,strings}

\submission{
  \vfill
  {\small\medskip\noindent{\bf Open Access} This chapter is licensed under the terms of the Creative Commons\break Attribution 4.0 International License (\url{http://creativecommons.org/licenses/by/4.0/}), which permits use, sharing, adaptation, distribution and reproduction in any medium or format, as long as you give appropriate credit to the original author(s) and the source, provide a link to the Creative Commons license and indicate if changes were made.}

  {\small \spaceskip .28em plus .1em minus .1em The images or other third party material in this chapter are included in the chapter's Creative Commons license, unless indicated otherwise in a credit line to the material.~If material is not included in the chapter's Creative Commons license and your intended\break use is not permitted by statutory regulation or exceeds the permitted use, you will need to obtain permission directly from the copyright holder.}

  \medskip\noindent\includegraphics{cc_by_4-0.eps}
}

\end{document}